\pdfoutput=1
\documentclass[11pt,twoside,a4paper,cmspaper,final,collab]{cms-tdr}
\def\svnVersion{61d324c}\def\svnDate{2022/05/06}\def\cmsCernNoTag{CERN-EP-2021-191}\def\cmsCernDate{\today}\def\cmsMessage{Published in Physical Review Letters as \href{http://dx.doi.org/10.1103/PhysRevLett.129.011801}{\doi{10.1103/PhysRevLett.129.011801}.}}
\begin{document}\cmsNoteHeader{EXO-18-014}

\renewcommand{\cmsCollabName}{The CMS and TOTEM Collaborations}
\renewcommand{\cmslogo}{\includegraphics[height=2.33cm]{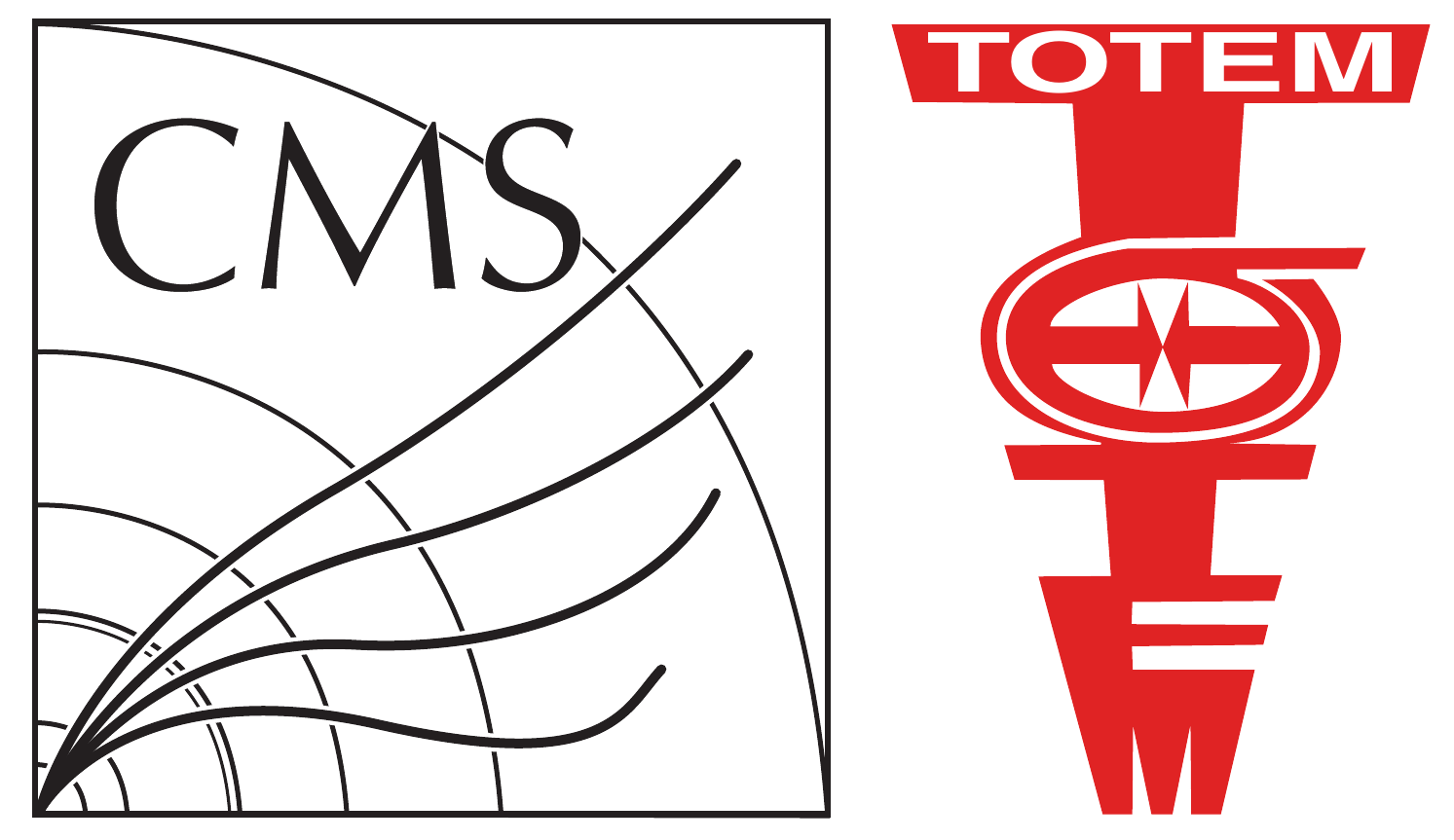}}
\renewcommand{\cmsTag}{CMS-\cmsNUMBER\\&TOTEM-2021-002\\}
\renewcommand{\appMsg}{See Appendices \ref{app:collab} and \ref{app:totem} for the list of collaboration members}

\title{First search for exclusive diphoton production at high mass with tagged protons in proton-proton collisions at \texorpdfstring{$\sqrt{s} = 13\TeV$}{sqrt(s) = 13 TeV}}
\author{The CMS and TOTEM Collaborations}
\date{\today}

\newcommand{\mpp}{\ensuremath{m_{\Pp\Pp}}\xspace}
\newcommand{\mgg}{\ensuremath{m_{\PGg\PGg}}\xspace}
\newcommand{\xigg}{\ensuremath{\xi_{\PGg\PGg}^\pm}\xspace}

\abstract{
A search for exclusive two-photon production via photon exchange in proton-proton collisions, $\Pp\Pp\to\Pp\PGg\PGg\Pp$ with intact protons, is presented. The data correspond to an integrated luminosity of 9.4\fbinv collected in 2016 using the CMS and TOTEM detectors at a center-of-mass energy of 13\TeV at the LHC. Events are selected with a diphoton invariant mass above 350\GeV and with both protons intact in the final state, to reduce backgrounds from strong interactions. The events of interest are those where the invariant mass and rapidity calculated from the momentum losses of the forward-moving protons matches the mass and rapidity of the central, two-photon system. No events are found that satisfy this condition. Interpreting this result in an effective dimension-8 extension of the standard model, the first limits are set on the two anomalous four-photon coupling parameters. If the other parameter is constrained to its standard model value, the limits at 95\% \CL are $\abs{\zeta_1}<2.9\ten{-13}\GeV^{-4}$ and $\abs{\zeta_2}<6.0\ten{-13}\GeV^{-4}$.
}

\hypersetup{
pdfauthor={CMS Collaboration},
pdftitle={First search for exclusive diphoton production at high mass with tagged protons in proton-proton collisions at 13 TeV at the LHC},
pdfsubject={CMS, TOTEM},
pdfkeywords={CMS, TOTEM, diphoton, tagged protons}}

\maketitle
In the classical theory of electrodynamics, photons (\PGg), having neither mass nor charge, lack self interactions.
However, because of the characteristics of the vacuum, photons with sufficient energy may fluctuate into charged particle-antiparticle pairs, thus giving rise to photon-photon interactions.
When two photons interact in this way through an intermediate charged particle loop to create two different outgoing photons, the process is known as light-by-light (LbL) scattering.
Evidence for this process has been sought in laboratory experiments for decades \cite{Adler:1971wn,Jarlskog:1974tx,Lee:1997rta,Bregant:2008yb}, and has been studied indirectly by the measurement of the anomalous magnetic moments of the electron \cite{VanDyck:1987ay} and the muon \cite{Bennett:2006fi,Abi:2021gix}.

By exploiting the large photon fluxes produced by ultrarelativistic ion beams at the LHC, as proposed in Ref. \cite{dEnterria:2013zqi}, the ATLAS and CMS experiments recently reported measurements of LbL scattering \cite{ATLAS:2019juy,Sirunyan:2018fhl,Aaboud:2017bwk} through exclusive diphoton production in ultraperipheral lead-lead collisions \cite{Baltz:2007kq}.
Analyses based on heavy ion collision data find results consistent with the standard model (SM) expectations.
However, these analyses probe the production of LbL candidates in the diphoton mass ($\mgg$) range of a few \GeV.
Complementary to these searches, an exclusive $\mgg$ spectrum search starting from 350\GeV is performed in this Letter for the first time.

The LbL scattering process, which can be studied at the electroweak energy scale and above in proton-proton ($\Pp\Pp$) collisions at the LHC, is of great interest because of its sensitivity to many extensions of the SM \cite{dEnterria:2013zqi,Chapon:2009hh,Sahin:2009gq,Gupta:2011be,Epele:2012jn,Fichet:2013gsa,Sun:2014qba,Fichet:2016clq}.
Some of these can be described by a purely effective extension of the SM Lagrangian using charge conjugation conserving operators, leading to a dimension-8 term for the four-photon coupling.
This term contains the electromagnetic field tensor, $F$, and the two parameters $\zeta_{1,2}$:
\begin{equation}
  L_8^{\PGg\PGg\PGg\PGg}=\zeta_1 F_{\mu\nu}F^{\mu\nu}F_{\rho\sigma}F^{\rho\sigma}+\zeta_2 F_{\mu\nu}F^{\mu\rho}F_{\rho\sigma}F^{\sigma\nu}.
\end{equation}
The contribution from the anomalous four-photon coupling would dominate the LbL cross section at high masses, as compared to the SM contribution \cite{Fichet:2014uka}, for even the small values of the coupling constants of order $10^{-14}\GeV^{-4}$ to which the analysis is sensitive.
Observing two intact protons and two photons in the CMS detector with the present accumulated luminosity would be a clear signal for beyond-the-SM physics.
A similar approach was used in Refs. \cite{Chatrchyan:2013akv,Khachatryan:2016mud,Aaboud:2016dkv} for the $\PGg\PGg\PWp\PWm$ quartic coupling.

\begin{figure}[ht]
  \centering
  \includegraphics[width=.35\textwidth]{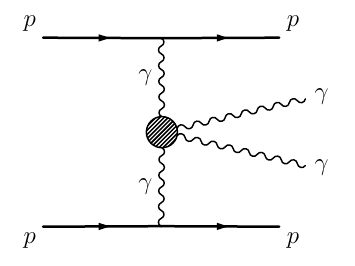}
  \caption{
    The process for diphoton production via photon exchange with intact protons in the final state.
    Several couplings may enter the four-photon shaded area such as a loop (box) of charged fermions or bosons.
    The model can be extended with intermediate interactions of new physics objects, such as a loop of a heavy charged particle or an $s$-channel process producing a scalar axion-like resonance that decays into two photons.}
  \label{fig:diag_gggg}
\end{figure}

In $\Pp\Pp$ collisions, LbL scattering (pictured in Fig. \ref{fig:diag_gggg}) can be identified through the measurement of two exclusively produced photons and two intact protons detected in very forward detectors along both beam directions.
In this Letter, a search for this process is performed in $\Pp\Pp$ collisions at a center-of-mass energy of 13\TeV using data collected with the CMS and TOTEM detectors in 2016, corresponding to an integrated luminosity of 9.4\fbinv.
Tabulated results are provided in the HEPData record for this analysis~\cite{hepdata}.

The basic feature of the CMS detector is the 3.8 T magnetic field produced by a superconducting solenoid.
Within this field are located a silicon pixel and strip tracker with coverage in pseudorapidity up to $|\eta|=2.5$, surrounded by a lead tungstate crystal electromagnetic calorimeter (ECAL), and a brass and scintillator hadron calorimeter (HCAL) directly outside the ECAL, each composed of a barrel and two endcap sections.
The ECAL consists of about 76\,000 PbWO${}_{4}$ crystals, each with a transverse dimension approximately matching the Moli\`{e}re radius of the material.
This radius roughly corresponds to a $\Delta\eta{\times}\Delta\phi$ granularity (where $\phi$ is azimuthal angle in radians) of $0.0174{\times}0.0174$ in the barrel, and extending up to $0.05{\times}0.05$ in both endcaps.
The muon detection system consists of three types of gas-ionization detectors located in the steel flux-return yoke of the solenoid.
Events are selected online and stored at a maximal rate of about 1 kHz using a two-tier triggering system \cite{Khachatryan:2016bia}.
A more detailed description of the apparatus, with a definition of the coordinate system used and the relevant kinematic variables, can be found in Ref. \cite{Chatrchyan:2008aa}.

The CMS-TOTEM precision proton spectrometer (CT-PPS) \cite{Albrow:2014lrm} is an array of movable, near-beam ``Roman pot" (RPs) devices containing tracking and timing detectors located with their inner edge at a distance of 1.5\mm from the nominal axis of the LHC beam, whose transverse width is about 0.1\mm.
The detectors are used to reconstruct the flight path and time of arrival of protons coming from the interaction point (IP) to a point 210 m down the LHC beamline.
In this study, two tracking stations per side, or ``arm'', of CMS are used.
These tracking stations provide a measurement of the proton trajectories with respect to the beam position.
Knowledge of the magnetic fields traversed by the proton from the IP to the RPs allows for the reconstruction of its fractional momentum loss $\xi=\Delta p/p\sim x/D_x$ with respect to the momentum of the incident proton, with $x$ being the horizontal displacement of the scattered proton, and $D_x$ being the horizontal dispersion of the beam.
In the 2016 data taking configuration, the CT-PPS tracking component consisted of silicon strip detectors, with acceptance for $\xi$ of 2.4--15\% and 3.7--15\%, respectively, for the two arms of the spectrometer.
The upper limit on this range reflects the acceptance expected from the nominal position of the physical apertures of optical elements in the beamline lattice in front of the RPs.
The techniques used for the alignment and calibration of the apparatus are detailed in Refs. \cite{Kaspar:2256296,Nemes:2256433}.
Combining the uncertainties in the alignment and spatial resolution of the RPs, the beam transverse size and angular spread, and the horizontal dispersion, a total relative uncertainty between 6--10\% is estimated for $\xi$ over the range of the detector acceptance.
The performance of CT-PPS and its potential for high-mass exclusive measurements were validated by the observation of proton-tagged $\PGg\PGg$ collisions in 2016 \cite{CMS:2018uvs}.
The CT-PPS silicon strips, by design, can only reconstruct one proton at a time.
This feature leads to a failure of the event reconstruction when multiple candidates are observed in the same RP for the same bunch crossing, leading to an inefficiency of 30\% or less in both arms of CT-PPS for the entire data taking period considered in this study.
Additionally, the acceptance is restricted to the regions of the silicon strips where the radiation-induced inefficiency remains below 10\%.
This region, asymmetric in proton momentum losses, corresponds to $0.07<\xi^{-}<0.111$ and $0.07<\xi^{+}<0.138$.

Events are required to pass a trigger that selects a pair of photon candidates, with each photon having a transverse momentum $\pt > 60\GeV$ each and a ratio of energy deposits in the HCAL to ECAL less than 0.15.
This trigger was designed for, and used in, the CMS inclusive high-mass diphoton searches \cite{Khachatryan:2016yec}.
In the case of an elastic, photon-induced process, the photon pair is expected to have different kinematic properties compared to inclusive processes where the photons are produced together with other particles.
In particular, the back-to-back momentum balance in the transverse plane between the two exclusive photons is used to select the exclusive two-photon events.

A boosted decision tree discriminant is used for the photon identification, following the procedure introduced in Refs. \cite{CMS:2020uim,Khachatryan:2016yec}.
Since the reconstruction algorithms in the ECAL do not make assumptions as to whether the energy deposits are from a photon or an electron, photon reconstruction can be validated using $\PZ\to\EE$ events \cite{Khachatryan:2015iwa}.
For this analysis, electrons are vetoed by the presence of hits in the central tracker that are inconsistent with a converted photon.

Several sources of nonexclusive backgrounds are considered for this search.
The leading inclusive $\PGg\PGg$ background as well as the $\PW\PGg$ and $\PZ\PGg$ subleading backgrounds are simulated by \MGvATNLO2.2.2 \cite{Alwall:2014hca} at next-to-leading order (NLO) precision with NNPDF3.0 parton distribution functions (PDFs) at next-to-NLO precision \cite{Ball:2014uwa}.
Other subleading backgrounds, namely photon-enriched quantum chromodynamics (QCD) processes, photon-enriched inclusive $\PGg+\text{jet}$, and inclusive $\ttbar$ processes, are generated at leading order (LO) by \PYTHIA 8.205 \cite{Sjostrand:2014zea} with NNPDF3.0 PDFs at LO precision.
The exclusive SM LbL process contribution is expected to be negligible at a mass range above 350\GeV \cite{Fichet:2014uka}.
Less than 0.02 events are expected for the luminosity used in this study.
It is considered as a background and is simulated using the Forward Physics Monte Carlo (\textsc{Fpmc}) program~\cite{Boonekamp:2011ky} based on the description in Ref. \cite{Fichet:2013gsa}.
All samples considered in this search are processed with a \GEANTfour \cite{Agostinelli:2002hh} simulation of the CMS central detector.

The selection is performed in the few steps detailed below, which are summarized in Table \ref{tab:selection}.
A preselection of events requires each photon to have $\pt>75\GeV$ to ensure a fully efficient trigger, $\abs{\eta}<2.5$ with a veto on the ECAL barrel-endcap transition region ($1.4442 < \abs{\eta} < 1.5660$), and $\mgg > 350\GeV$.
The minimum diphoton mass corresponds to the minimum measurable proton momentum loss detectable by the CT-PPS.

For events passing the preselection criteria, an elastic selection region is constructed based on the expected back-to-back emission in the transverse plane of two final-state photons from exclusive elastic processes.
The diphoton acoplanarity $a\equiv 1-\abs{\Delta\phi_{\PGg\PGg}/\pi}$, where $\Delta\phi_{\PGg\PGg}$ is the azimuthal separation of the two photons, is then used as a discriminating variable, and diphoton candidates are selected with $a < 0.005$.
To isolate photons in the ECAL, a lower threshold of 0.94 is set on the $\RNINE$ variable, computed as the ratio between the energy in a $3{\times}3$ area of crystals to the energy in a $5{\times}5$ area of crystals, centered on the most energetic crystal of the photon energy deposit \cite{Khachatryan:2015iwa}.
Using the criteria from the elastic selection based on the kinematic features of the two-photon system, which allows a selection from the central detector only, a tighter search can be defined with the detection of scattered protons for elastic production.
It requires both $\xi_{\PGg\PGg}^{-}$ and $\xi_{\PGg\PGg}^{+}$ to be compatible with the CT-PPS proton $\xi$ acceptance quoted earlier, where $\xigg = (\pt^{\PGg_1}\re^{\pm\eta^{\PGg_1}}+\pt^{\PGg_2}\re^{\pm\eta^{\PGg_2}})/\sqrt{s}$ is calculated from the two-photon system kinematic variables only and the $+$ and $-$ denote the positive and negative z sides of CMS, respectively.
Because of the radiation damage to the detector regions closest to the beam, the track reconstruction efficiency varies with $x$ (and hence $\xi$), and decreases over time as radiation is accumulated.
For that reason, the signal search region for this analysis is defined by requiring the two $\xigg$ values to pass a tighter selection corresponding to the most efficient area of the CT-PPS detectors.
In this region, equivalently noted as ``tight $\xigg$'' in this study, the tracking efficiency in each RP is at least 90\%.

Sensitivity to the LbL signal is enhanced by measuring the resulting final-state protons.
In exclusive events, where the protons remain intact, momentum loss from the protons is related to the invariant mass of the diphoton system.
Signal candidates are selected by requiring, in addition to the selection criteria defined above, a kinematic matching between the two systems, of the forward protons and the central photons, thus imposing conservation of momentum.
It has been shown that matching mass and rapidity of the diphoton system and the scattered protons on an event-by-event basis significantly reduces the contribution of inclusive backgrounds \cite{Fichet:2014uka}.
In fact, the large majority of such events come from the coincidence of an inclusively-produced diphoton event with pileup protons from unrelated events.
The kinematic matching ensures that the two systems originate from the same $\Pp\Pp$ interaction.
The kinematic variables of an opposite-arm, two-proton system are converted into missing mass and rapidity of the central system through $\mpp=\sqrt{\smash[b]{s\xi^+ \xi^-}}$, and $y_{\Pp\Pp}=(1/2)\log(\xi^+/\xi^-)$, where $\xi^+$ and $\xi^-$ correspond to the $\xi$ of protons on the positive z and negative z sides of CT-PPS, respectively.
In the case of exclusive diphoton production, both systems are correlated through $\mpp=\mgg$ and $y_{\Pp\Pp}=y_{\PGg\PGg}$.
The resolution of the diphoton mass as deducted from uncertainties in the photons' momenta is 2.0\%.
For the two-proton system, a diphoton mass resolution of 5.5--8.4\% is expected from the proton fractional momentum loss uncertainties.
Equivalently, the central two-photon rapidity resolution is 7.4\%, while the uncertainty in $y_{\Pp\Pp}$, the forward two-proton rapidity, is bounded between 0.05--0.09 in absolute value.
In this search, a 2$\sigma$ window is used in matching the difference, both in mass and rapidity, between the central and the two-proton systems following the tight $\xigg$ selection defined above; here $\sigma$ indicates the combined resolution of the two systems.

One nonelastic background control region is used in the analysis.
This control region satisfies the preselection criteria and is given a high efficiency for inclusive diphoton events with the requirement that photons have $\pt>200\GeV$ and $a>0.025$, with a looser $\RNINE>0.85$ criteria to ensure a high yield of the selection.
The acoplanarity selection suppresses exclusive production, which occurs at small acoplanarity values.
A summary of the selection at different stages is shown in Table \ref{tab:selection}.

\begin{table*}[htb]
  \centering
  \caption{Description of the selection stages applied on the photon and proton kinematic variables, and used in this study.}
  \begin{tabular}{ll}
  \hline
  Stage & Selection on photon variables\\
  \hline\\[-1em]
  Preselection & $\pt^{\PGg}>75\GeV$, $\abs{\eta_{\PGg}}<2.5$ with ECAL geometrical veto, $\RNINE>0.85$, $\mgg>350\GeV$\\[.5em]
  Nonelastic & Preselection, $\pt^{\PGg}>200\GeV$, $a>0.025$\\[.5em]
  Elastic & Preselection, $\RNINE>0.94$, $a<0.005$\\
  Accept. $\xigg$ & Elastic selection, $0.024<\xi_{\PGg\PGg}^{-}<0.15$, $0.037<\xi_{\PGg\PGg}^{+}<0.15$\\
  Tight $\xigg$ & Elastic selection, $0.07<\xi_{\PGg\PGg}^{-}<0.111$, $0.07<\xi_{\PGg\PGg}^{+}<0.138$\\
  2$\sigma$ match. & Tight $\xigg$ selection, 2$\sigma$ matching of $m_{\PGg\PGg}$ and $y_{\PGg\PGg}$ with $m_{\Pp\Pp}$ and $y_{\Pp\Pp}$\\[.25em]
  \hline
  \end{tabular}
  \label{tab:selection}
\end{table*}

The normalization and shape of the simulated background contributions in the signal search region are checked with the nonelastic-enriched sample introduced above.
A slight deficit (9.9\%) observed for the simulation of the nonelastic-enriched region is addressed by rescaling the dominant inclusive background to achieve an agreement between data and simulation.
This background yield correction is applied to each of the selection regions.

\begin{figure}[htb]
  \centering
  \includegraphics[width=.5\textwidth]{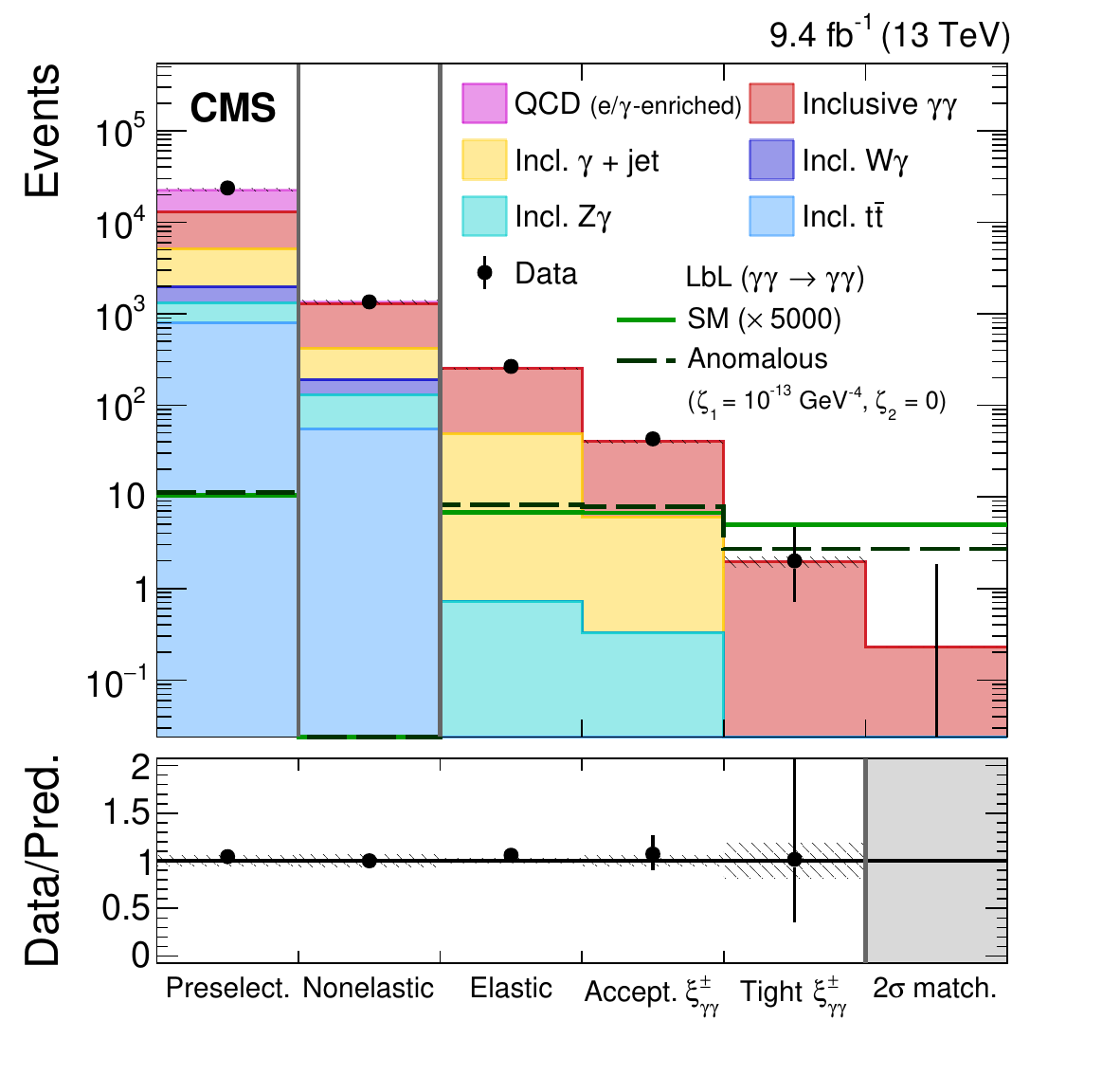}
  \caption{Numbers of simulated and observed events for the various selection regions described in the text.
The shaded bands show the statistical uncertainties in the simulated backgrounds added in quadrature.
All selection regions are sequential from left to right, with the exception of the nonelastic region used in the background yield correction, thus with a data-to-prediction ratio constrained to unity.
Shown also are the theoretical contributions to the LbL signal from the SM and from an anomalous source whose yield at the parameters given is close to the sensitivity reached in this study.}
  \label{fig:cutflow}
\end{figure}

\begin{figure}[htb]
  \centering
  \includegraphics[width=.5\textwidth]{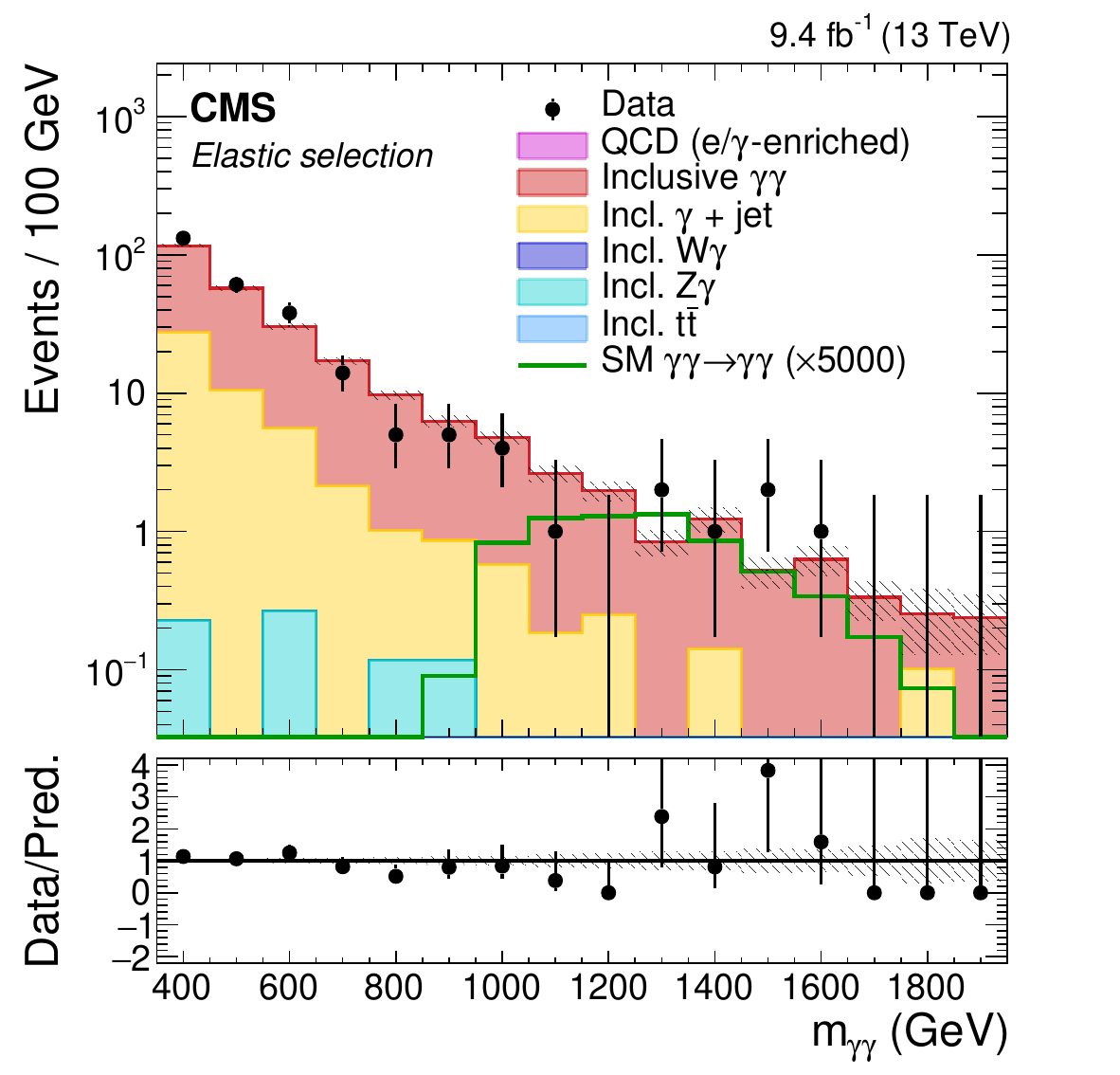}
  \caption{
    Invariant mass distribution of the diphoton pairs for the elastic selection region with events satisfying $1-\abs{\Delta\phi_{\PGg\PGg}/\pi}<0.005$ as described in the text, for data (dots) and MC simulations (histograms).
    The hatched bands indicate the statistical uncertainties in simulated samples added in quadrature.
  }
  \label{fig:kinematics}
\end{figure}

The data and simulation events falling within the different selection regions described above and given in Table \ref{tab:selection} can be seen in Fig. \ref{fig:cutflow}.
For all selection regions used in this study, the data are found to be consistent with the background prediction within statistical uncertainties.
A total of 266 diphoton candidates are found in the elastic region to be compared with the expectation of $263.1\pm 4.1$ \stat.
The resulting $\mgg$ spectrum of events passing the elastic selection can be seen in Fig. \ref{fig:kinematics}.

As shown in Fig. \ref{fig:cutflow}, only two events remain in the tight $\xigg$ signal search region with an expected background of $2.1^{+1.0}_{-0.7}$ \stat when no kinematic matching criteria between the two-photon and two-proton systems are applied.
Neither of these events contains a pair of forward proton tracks, and thus there are no events in the $2\sigma$ match region.

Background contributions are estimated following the procedure described in Ref. \cite{CMS:2018uvs}, where it is assumed that inclusive background processes involve a full decorrelation between central two-photon and forward two-proton systems.
Pseudo-events are formed by combining diphoton kinematic distributions sampled from a template with the two-proton system variables randomly selected from real data events within the period of interest.
The diphoton kinematic variables are sampled from an exponential fit to the two $\xigg$ spectra of events passing the nonelastic background-enriched selection defined above.
In order to match the observed yield in data for the tight $\xigg$ selection and after checking that it does not show any kinematic dependence on the invariant mass, transverse momentum, and acoplanarity of the photon pair, or on the basic kinematic variables of single photons, a constant scaling factor is applied to the leading inclusive $\PGg\PGg$ background.
Using this method, the predicted number of events having an elastic diphoton pair in association with a pair of protons observed within the range where the proton detectors have a radiation inefficiency less than 10\% is evaluated as $0.83^{+0.28}_{-0.15}$ \stat events.
This prediction is without the requirement of any kinematic matching of the diphoton and proton systems.
When constraining both the systems to the $2\sigma$ and $3\sigma$ matching windows, the background predictions are respectively $0.23^{+0.08}_{-0.04}\stat$ and $0.43^{+0.14}_{-0.08}\stat$ events.

Several sources of systematic uncertainties affecting the signal are considered.
The uncertainty of 37\% in the yield correction for the inclusive background selection estimated from the nonelastic-enriched selection as described above.
This uncertainty is evaluated by varying the fit parameters within one standard deviation.
There is an additional 33\% uncertainty in the background estimate, evaluated by computing this yield after varying the exponential fit parameters within one standard deviation.
The variation on the functional form of the fit was also studied and was found to be well within the uncertainty quoted above.
Subleading sources of uncertainties include radiation damage and tracking efficiency of the RPs (13\%) \cite{CMS-DP-2020-047}, and the luminosity measurement (2.5\%) \cite{CMS:2021xjt}.
Additionally, a signal cross section uncertainty of 10\% is assumed to account for the rapidity gap survival probability in the high invariant mass region \cite{Harland-Lang:2015cta}.
The rapidity gap survival probability \cite{Harland-Lang:2016apc} expresses the fraction of events in which no additional soft interactions occur between the two colliding protons, producing extra final-state particles and modifying the topology of exclusive events.

No diphoton candidates with exclusive kinematic features are observed in either of the $2\sigma$ and 3$\sigma$ mass and rapidity matching windows, thus no observation of a central exclusive production of a photon pair with associated scattered proton tracks can be reported for the sample considered in this study.

Using a profile likelihood ratio as a test statistic \cite{Cowan:2010js}, systematic uncertainties as nuisance parameters with a log-normal prior, and the background yields obtained above, the 95\% confidence level (\CL) \cite{Junk:1999kv,Read:2002hq} observed frequentist upper limit of 4.4\unit{fb} is obtained for the LbL cross section within the fiducial region.
This region is defined at the generator level in terms of the single-photon and diphoton selections described previously, with additional asymmetric selection criteria for the two arms of the spectrometer corresponding to the region with less than 10\% inefficiency from the CT-PPS strips radiation damage and within beamline apertures.
The SM LbL process has a combined two-photon and two-proton tagging signal efficiency of (6.7 $\pm$ 0.9)\%.
The dominant fraction of the uncertainty in this efficiency arises from the uncertainties in estimating radiation damage and reconstruction of multiple protons in the RPs, followed by a contribution driven by the limited number of SM LbL MC events satisfying the selection criteria.

\begin{figure}[htb]
\centering
  \includegraphics[width=.5\textwidth]{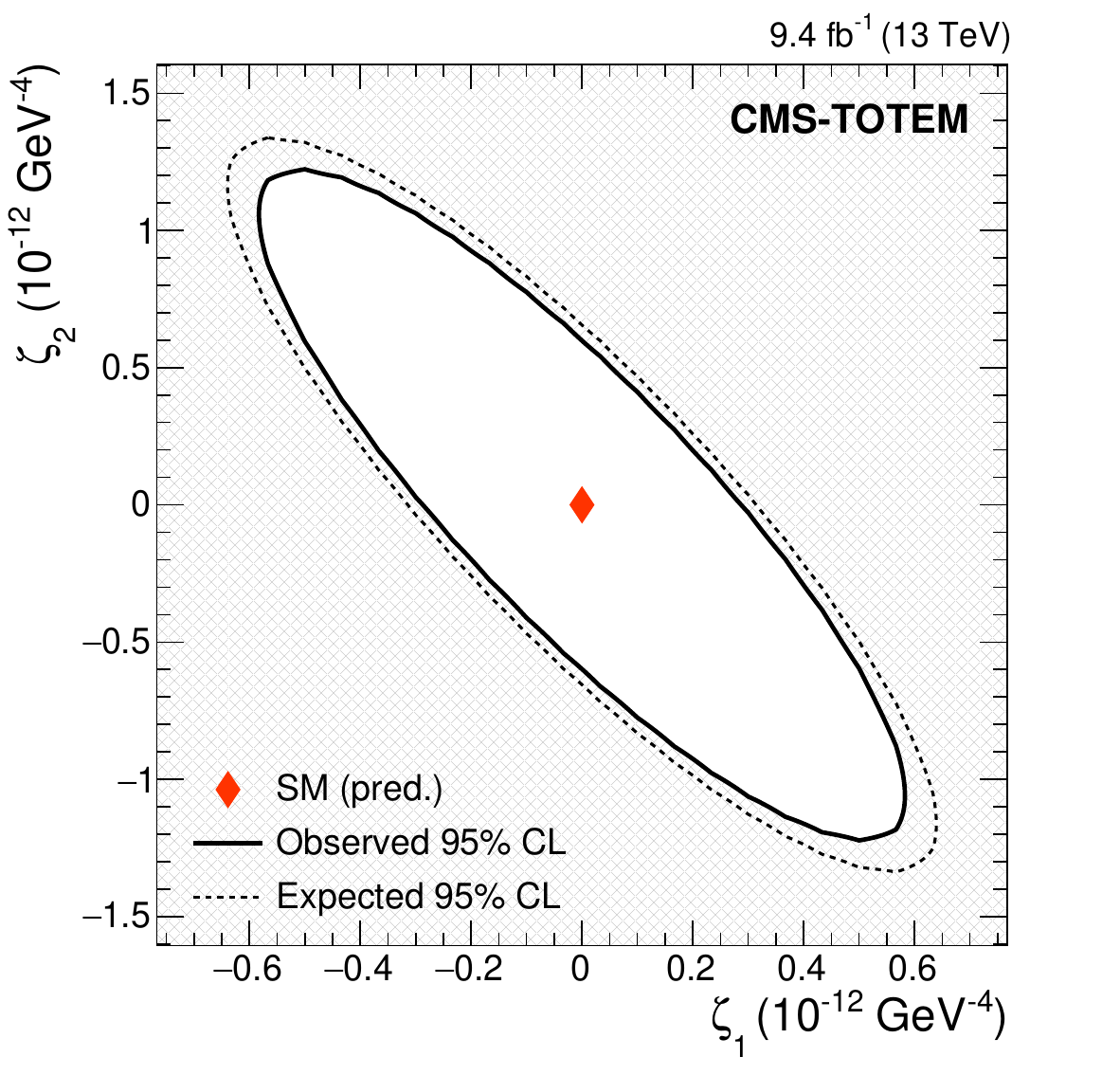}
  \caption{Two-dimensional limits on the anomalous four-photon couplings, derived from the observed upper limit on the diphoton production cross section. The shaded area depicts the excluded values of the coupling parameters $\zeta_1$ and $\zeta_2$.}
  \label{fig:scan_anom}
\end{figure}

For the anomalous quartic gauge coupling extension of the SM introduced earlier, an observed upper limit of 2.1\unit{fb} can be compared with the expected limit of 2.5\unit{fb} using the background-only hypothesis.
This upper limit is used to place the first limits on the four-photon anomalous quartic gauge couplings.
It was studied that the signal efficiency does not vary strongly over a wide range of the couplings parameters $\zeta_1$ and $\zeta_2$ in the search region.
It is evaluated at a combined (14.5 $\pm$ 1.9)\% signal efficiency, significantly higher than for the SM LbL process.
Figure \ref{fig:scan_anom} shows the region of the parameter phase space where the corresponding cross section is excluded by this measurement.
Consequently, when one of the model parameters is assumed to be null, the other is limited to
\begin{linenomath}
\begin{equation*}
\begin{aligned}
\abs{\zeta_1} < 2.9 \ten{-13}\GeV^{-4} & (\zeta_2 = 0),\\
\abs{\zeta_2} < 6.0 \ten{-13}\GeV^{-4} & (\zeta_1 = 0).
\end{aligned}
\end{equation*}
\end{linenomath}

To summarize, the CMS-TOTEM precision proton spectrometer has proven the feasibility of continuously operating a near-beam proton spectrometer at a high-luminosity hadron collider.
The first search for the $\PGg\PGg\to\PGg\PGg$ process with forward proton tags is presented.
The search uses an integrated luminosity of 9.4\fbinv of proton-proton collisions collected at a 13\TeV center-of-mass energy at the LHC during 2016.
No events are observed with a pair of proton tracks compatible with the diphoton kinematic properties with an expected background of 0.23 and 0.43 events for the 2 and 3 standard deviations windows, respectively.
This provides the first limit for the standard model light-by-light production cross section at a scale of hundreds of \GeV, and places limits on anomalous couplings for the four-photon interaction based on an effective field theory extension of the standard model.

\begin{acknowledgments}
We congratulate our colleagues in the CERN accelerator departments for the excellent performance of the LHC and thank the technical and administrative staffs at CERN and at other CMS and TOTEM institutes for their contributions to the success of the common CMS-TOTEM effort. In addition, we gratefully acknowledge the computing centers and personnel of the Worldwide LHC Computing Grid and other centers for delivering so effectively the computing infrastructure essential to our analyses. Finally, we acknowledge the enduring support for the construction and operation of the LHC, the CMS and TOTEM detectors, and the supporting computing infrastructure provided by the following funding agencies: BMBWF and FWF (Austria); FNRS and FWO (Belgium); CNPq, CAPES, FAPERJ, FAPERGS, and FAPESP (Brazil); MES and BNSF (Bulgaria); CERN; CAS, MoST, and NSFC (China); MINCIENCIAS (Colombia); MSES and CSF (Croatia); RIF (Cyprus); SENESCYT (Ecuador); MoER, ERC PUT and ERDF (Estonia); Academy of Finland, Finnish Academy of Science and Letters (The Vilho Yrj\"{o} and Kalle V\"{a}is\"{a}l\"{a} Fund), MEC, Magnus Ehrnrooth Foundation, HIP, and Waldemar von Frenckell Foundation (Finland); CEA and CNRS/IN2P3 (France); BMBF, DFG, and HGF (Germany); GSRI (Greece); the Circles of Knowledge Club, NKFIA (Hungary); DAE and DST (India); IPM (Iran); SFI (Ireland); INFN (Italy); MSIP and NRF (Republic of Korea); MES (Latvia); LAS (Lithuania); MOE and UM (Malaysia); BUAP, CINVESTAV, CONACYT, LNS, SEP, and UASLP-FAI (Mexico); MOS (Montenegro); MBIE (New Zealand); PAEC (Pakistan); MSHE and NSC (Poland); FCT (Portugal); JINR (Dubna); MON, RosAtom, RAS, RFBR, and NRC KI (Russia); MESTD (Serbia); SEIDI, CPAN, PCTI, and FEDER (Spain); MOSTR (Sri Lanka); Swiss Funding Agencies (Switzerland); MST (Taipei); ThEPCenter, IPST, STAR, and NSTDA (Thailand); TUBITAK and TAEK (Turkey); NASU (Ukraine); STFC (United Kingdom); DOE and NSF (USA).
\end{acknowledgments}

\bibliography{auto_generated}
\cleardoublepage \appendix\section{The CMS Collaboration \label{app:collab}}\begin{sloppypar}\hyphenpenalty=5000\widowpenalty=500\clubpenalty=5000\vskip\cmsinstskip
\textbf{Yerevan Physics Institute, Yerevan, Armenia}\\*[0pt]
A.~Tumasyan
\vskip\cmsinstskip
\textbf{Institut f\"{u}r Hochenergiephysik, Vienna, Austria}\\*[0pt]
W.~Adam, T.~Bergauer, M.~Dragicevic, J.~Er\"{o}, A.~Escalante~Del~Valle, R.~Fr\"{u}hwirth\cmsAuthorMark{1}, M.~Jeitler\cmsAuthorMark{1}, N.~Krammer, L.~Lechner, D.~Liko, I.~Mikulec, F.M.~Pitters, N.~Rad, J.~Schieck\cmsAuthorMark{1}, R.~Sch\"{o}fbeck, M.~Spanring, S.~Templ, W.~Waltenberger, C.-E.~Wulz\cmsAuthorMark{1}, M.~Zarucki
% ...
\vskip\cmsinstskip
% ...
1:  Also at TU Wien, Wien, Austria\\
% ...

\section{The TOTEM Collaboration\label{app:totem}}
\newcommand{\AddAuthor}[2]{#1$^{#2}$,\ }
\newcommand\AddAuthorLast[2]{#1$^{#2}$}
\noindent
\AddAuthor{G.~Antchev}{a}
\AddAuthor{P.~Aspell}{9}
\AddAuthor{I.~Atanassov}{a}
\AddAuthor{V.~Avati}{7,9}
\AddAuthor{J.~Baechler}{9}
\AddAuthor{C.~Baldenegro~Barrera}{11}
\AddAuthor{V.~Berardi}{4a,4b}
\AddAuthor{M.~Berretti}{2a}
\AddAuthor{V.~Borchsh}{8}
\AddAuthor{E.~Bossini}{6b,9}
\AddAuthor{U.~Bottigli}{6c}
\AddAuthor{M.~Bozzo}{5a,5b}
\AddAuthor{H.~Burkhardt}{9}
\AddAuthor{F.S.~Cafagna}{4a}
\AddAuthor{M.G.~Catanesi}{4a}
\AddAuthor{M.~Csan\'{a}d}{3a,b}
\AddAuthor{T.~Cs\"{o}rg\H{o}}{3a,3b}
\AddAuthor{M.~Deile}{9}
\AddAuthor{F.~De~Leonardis}{4a,4c}
\AddAuthor{M.~Doubek}{1c}
\AddAuthor{D.~Druzhkin}{8,9}
\AddAuthor{K.~Eggert}{10}
\AddAuthor{V.~Eremin}{d}
\AddAuthor{A.~Fiergolski}{9}
\AddAuthor{L.~Forthomme}{2a,2b}
\AddAuthor{F.~Garcia}{2a}
\AddAuthor{V.~Georgiev}{1a}
\AddAuthor{S.~Giani}{9}
\AddAuthor{L.~Grzanka}{7}
\AddAuthor{J.~Hammerbauer}{1a}
\AddAuthor{T.~Isidori}{11}
\AddAuthor{V.~Ivanchenko}{8}
\AddAuthor{M.~Janda}{1c}
\AddAuthor{A.~Karev}{9}
\AddAuthor{J.~Ka\v{s}par}{1b,9}
\AddAuthor{B.~Kaynak}{e}
\AddAuthor{J.~Kopal}{9}
\AddAuthor{V.~Kundr\'{a}t}{1b}
\AddAuthor{S.~Lami}{6a}
\AddAuthor{R.~Linhart}{1a}
\AddAuthor{C.~Lindsey}{11}
\AddAuthor{M.V.~Lokaj\'{i}\v{c}ek$^{\textrm{\dag,}}$}{1b}
\AddAuthor{L.~Losurdo}{6c}
\AddAuthor{F.~Lucas~Rodr\'{i}guez}{9}
\AddAuthor{M.~Macr\'{i}}{5a}
\AddAuthor{M.~Malawski}{7}
\AddAuthor{N.~Minafra}{11}
\AddAuthor{S.~Minutoli}{5a}
\AddAuthor{T.~Naaranoja}{2a,2b}
\AddAuthor{F.~Nemes}{3a,9}
\AddAuthor{H.~Niewiadomski}{10}
\AddAuthor{T.~Nov\'{a}k}{3b}
\AddAuthor{E.~Oliveri}{9}
\AddAuthor{F.~Oljemark}{2a,2b}
\AddAuthor{M.~Oriunno}{f}
\AddAuthor{K.~\"{O}sterberg}{2a,2b}
\AddAuthor{P.~Palazzi}{9}
\AddAuthor{V.~Passaro}{4a,4c}
\AddAuthor{Z.~Peroutka}{1a}
\AddAuthor{J.~Proch\'{a}zka}{1b}
\AddAuthor{M.~Quinto}{4a,4b}
\AddAuthor{E.~Radermacher}{9}
\AddAuthor{E.~Radicioni}{4a}
\AddAuthor{F.~Ravotti}{9}
\AddAuthor{C.~Royon}{11}
\AddAuthor{G.~Ruggiero}{9}
\AddAuthor{H.~Saarikko}{2a,2b}
\AddAuthor{V.D.~Samoylenko}{c}
\AddAuthor{A.~Scribano}{6a}
\AddAuthor{J.~\v{S}irok\'{y}}{1a}
\AddAuthor{J.~Smajek}{9}
\AddAuthor{W.~Snoeys}{9}
\AddAuthor{R.~Stefanovitch}{9}
\AddAuthor{J.~Sziklai}{3a}
\AddAuthor{C.~Taylor}{10}
\AddAuthor{E.~Tcherniaev}{8}
\AddAuthor{N.~Turini}{6c}
\AddAuthor{O.~Urban}{1a}
\AddAuthor{V.~Vacek}{1c}
\AddAuthor{O.~Vavroch}{1a}
\AddAuthor{J.~Welti}{2a,2b}
\AddAuthor{J.~Williams}{11}
\AddAuthor{J.~Zich}{1a}
\AddAuthorLast{K.~Zielinski}{7}

\vskip 4pt plus 4pt
\let\thefootnote\relax
\newcommand{\AddInstitute}[2]{${}^{#1}$#2\\}
\newcommand{\AddExternalInstitute}[2]{\footnote{${}^{#1}$ #2}}
\noindent
\dag Deceased\\
\AddInstitute{1a}{University of West Bohemia, Pilsen, Czech Republic}
\AddInstitute{1b}{Institute of Physics of the Academy of Sciences of the Czech Republic, Prague, Czech Republic}
\AddInstitute{1c}{Czech Technical University, Prague, Czech Republic}
\AddInstitute{2a}{Helsinki Institute of Physics, University of Helsinki, Helsinki, Finland}
\AddInstitute{2b}{Department of Physics, University of Helsinki, Helsinki, Finland}
\AddInstitute{3a}{Wigner Research Centre for Physics, RMKI, Budapest, Hungary}
\AddInstitute{3b}{Eszterhazy Karoly University KRC, Gy\"{o}ngy\"{o}s, Hungary}
\AddInstitute{4a}{INFN Sezione di Bari, Bari, Italy}
\AddInstitute{4b}{Dipartimento Interateneo di Fisica di Bari, University of Bari, Bari, Italy}
\AddInstitute{4c}{Dipartimento di Ingegneria Elettrica e dell'Informazione --- Politecnico di Bari, Bari, Italy}
\AddInstitute{5a}{INFN Sezione di Genova, Genova, Italy}
\AddInstitute{5b}{Universit\`{a} degli Studi di Genova, Genova, Italy}
\AddInstitute{6a}{INFN Sezione di Pisa, Pisa, Italy}
\AddInstitute{6b}{Universit\`{a} degli Studi di Pisa, Pisa, Italy}
\AddInstitute{6c}{Universit\`{a} degli Studi di Siena and Gruppo Collegato INFN di Siena, Siena, Italy}
\AddInstitute{7}{Akademia G\'{o}rniczo-Hutnicza (AGH) University of Science and Technology, Krakow, Poland}
\AddInstitute{8}{Tomsk State University, Tomsk, Russia}
\AddInstitute{9}{CERN, Geneva, Switzerland}
\AddInstitute{10}{Case Western Reserve University, Department of Physics, Cleveland, Ohio, USA}
\AddInstitute{11}{The University of Kansas, Lawrence, Kansas, USA}
\AddExternalInstitute{a}{INRNE-BAS, Institute for Nuclear Research and Nuclear Energy, Bulgarian Academy of Sciences, Sofia, Bulgaria}
\AddExternalInstitute{b}{Department of Atomic Physics, E\"{o}tv\"{o}s Lor\'{a}nd University, Budapest, Hungary}
\AddExternalInstitute{c}{NRC 'Kurchatov Institute'--IHEP, Protvino, Russia}
\AddExternalInstitute{d}{Ioffe Physical Technical Institute, Russian Academy of Sciences, St. Petersburg, Russian Federation}
\AddExternalInstitute{e}{Istanbul University, Istanbul, Turkey}
\AddExternalInstitute{f}{SLAC, Stanford University, California, USA}\end{sloppypar}
\end{document}